\documentclass{article}%
\usepackage{amsfonts}
\usepackage{amsthm}
\usepackage{amsmath}%
\usepackage{amssymb}%
\usepackage{graphicx}

\newtheorem{theorem}{Theorem}

\begin{document}

\title{Berezin--Toeplitz quantization on Lie groups}
\author{Brian C. Hall}
\maketitle

\begin{abstract}
Let $K$ be a connected compact semisimple Lie group and $K_{\mathbb{C}}$ its
complexification. The generalized Segal--Bargmann space for $K_{\mathbb{C}},$
is a space of square-integrable holomorphic functions on $K_{\mathbb{C}},$
with respect to a $K$-invariant heat kernel measure. This space is connected
to the \textquotedblleft Schr\"{o}dinger\textquotedblright\ Hilbert space
$L^{2}(K)$ by a unitary map, the generalized Segal--Bargmann transform.

This paper considers certain natural operators on $L^{2}(K),$ namely
multiplication operators and differential operators, conjugated by the
generalized Segal--Bargmann transform. The main results show that the
resulting operators on the generalized Segal--Bargmann space can be
represented as Toeplitz operators. The symbols of these Toeplitz operators are
expressed in terms of a certain subelliptic heat kernel on $K_{\mathbb{C}}.$

I also examine some of the results from an infinite-dimensional point of view
based on the work of L. Gross and P. Malliavin.

\end{abstract}

Address: University of Notre Dame, Department of Mathematics, Notre Dame IN
46556-4618 USA

E-mail: bhall{@}nd.edu

Thanks: Supported in part by NSF Grant DMS-02000649


\section{Introduction}

The Berezin--Toeplitz quantization is a standard method of quantizing a
symplectic manifold $\mathcal{M}$ that admits a K\"{a}hler structure. In such
cases, the quantum Hilbert space is a space of square-integrable holomorphic
sections of an appropriate complex line bundle. Let $P$ denote the orthogonal
projection operator from the space of all square-integrable sections to the
holomorphic subspace. Then for any bounded measurable function $\phi,$ we can
construct the \textit{Toeplitz operator with symbol }$\phi,$ acting on the
space of holomorphic sections, as $T_{\phi}s=P(\phi s).$ That is, $T_{\phi}$
consists of multiplication by $\phi$ followed by projection back into the
holomorphic subspace.

The map sending $\phi$ to $T_{\phi}$ is called the \textit{Berezin--Toeplitz
quantization}, and it may be thought of as a generalization of
anti-Wick-ordered quantization. (See \cite{mexnotes} for discussion.) There is
a large literature devoted to this quantization scheme, including early works
such as \cite{Be,Ra,RCG}, continuing with specific examples in
\cite{KL,Co,BLU}, and then developing into a general theory in \cite{BMS,BPU},
to mention just a few examples.

In this paper, we will examine the case in which $\mathcal{M}$ is the
cotangent bundle $T^{\ast}(K)$ of a connected compact Lie group $K,$ which we
assume for simplicity to be semisimple. (In the torus case, the results are
essentially the same as in the $\mathbb{R}^{n}$ case, but the semisimple case
displays some interesting new phenomena.) There is a natural way to identify
$T^{\ast}(K)$ with the \textit{complexification }$K_{\mathbb{C}}$ of $K$ and
in this case, the relevant line bundle is holomorphically trivial. Thus, in
this case, the quantum Hilbert space is identified with $\mathcal{H}%
L^{2}(K_{\mathbb{C}},\nu_{t}),$ the space of square-integrable holomorphic
functions on $K_{\mathbb{C}}$ with respect to a certain measure $\nu
_{t}(g)~dg,$ where $dg$ is a Haar measure on $K_{\mathbb{C}}$ and $\nu_{t}$ is
a $K$-invariant heat kernel. Here $t$ is a positive parameter that plays the
role of Planck's constant.

Of course, since $T^{\ast}(K)$ is a cotangent bundle, there is another
commonly used method of quantizing it, namely, Schr\"{o}dinger-style
quantization in which the Hilbert space is $L^{2}(K)$ with respect to a Haar
measure. The goal of the present paper is to compare the two approaches to
quantization. There is a natural unitary map between $L^{2}(K)$ and
$\mathcal{H}L^{2}(K_{\mathbb{C}},\nu_{t}),$ called the (generalized)
Segal--Bargmann transform and introduced in \cite{H1}. The goal of the present
paper is to show that certain natural operators on $L^{2}(K),$ when conjugated
by the Segal--Bargmann transform, become Toeplitz operators on $\mathcal{H}%
L^{2}(K_{\mathbb{C}},\nu_{t}).$

In Section \ref{mult.sec}, we consider $M_{V},$ multiplication by $V$, acting
as an operator on $L^{2}(K).$ The operator $M_{V}$ should be thought of as the
Schr\"{o}dinger quantization of the function $V\circ\pi,$ where $\pi$ is the
projection from $T^{\ast}(K)$ to $K.$ That is to say, multiplication operators
are the Schr\"{o}dinger quantization of functions that are constant along the
fibers of the cotangent bundle.

If $V$ is very regular, then $C_{t}M_{V}C_{t}^{-1}$ can be expressed as a
Toeplitz operator with symbol $\phi_{V},$ which is computed as follows. We
first apply the \textit{backward} heat operator $e^{-t\Delta/4}$ to $V,$
obtaining $\tilde{V}:=e^{-t\Delta/4}V.$ (For this to make sense, $V$ must be
very regular.) In the case where $K$ is commutative, $\phi_{V}$ is given
simply by $\phi_{V}(xe^{iy})=\tilde{V}(x).$ This is essentially the same as
what one has in the $\mathbb{R}^{n}$ case. On the other hand, if $K$ is
semisimple, then
\[
\phi_{V}(g)=\frac{\int_{K}\mu_{t/2,t}(gx^{-1})\tilde{V}(x)~dx}{\nu_{t}%
(g)},\quad g\in K_{\mathbb{C}},
\]
where $\mu_{t/2,t}$ is the heat kernel associated to a certain
\textit{subelliptic} Laplace-type operator on $K_{\mathbb{C}}.$

Even in the $\mathbb{R}^{n}$ case, the formula for $\phi_{V}$ involves
applying the backward heat operator $e^{-t\Delta/4}$ to $V,$ so the assumption
that $V$ is very regular (in the domain of the backward heat operator) seems
unavoidable. This assumption reflects a sort of smoothing property of the
Berezin--Toeplitz quantization, namely that even very rough symbols give rise
to nice operators. For example, it is easy to have a highly singular
distributional symbol $\phi$ for which the associated Toeplitz operator
$T_{\phi}$ is bounded. It follows that the inverse operation, trying to
represent an operator on $\mathcal{H}L^{2}(K_{\mathbb{C}},\nu_{t})$ as a
Toeplitz operator, will be quite singular.

In Section \ref{diff.sec}, we consider differential operators on $K$, which we
do not assume are invariant under either the left- or right-action of $K.$
Differential operators of degree at most $N$ should be thought of as the
quantization of functions on $T^{\ast}(K)$ that are polynomials of degree at
most $N$ on each fiber. (Specifically, one may consider a generalization to
$K$ of the Weyl quantization, which indeed maps the space of functions that
are polynomials of degree at most $N$ on each fiber into the space of
differential operators of degree at most $N.$) A differential operator on $K$
can be represented as a linear combination of left-invariant differential
operators multiplied by functions on $K,$ that is, as linear combinations of
left-invariant differential operators composed with multiplication operators.
If $\alpha$ is a left-invariant differential operator and $V$ is a
sufficiently regular function, then $C_{t}M_{V}\alpha C_{t}^{-1}$ can be
represented as a Toeplitz operator with a certain symbol. The results of
Section \ref{diff.sec} are a substantial generalization of the results of
\cite{HL}, which treats only the invariant case.

Finally, in Section \ref{inf.sec}, we look at the results of Section
\ref{mult.sec} from an infinite-dimensional point of view. The work of L.
Gross and P. Malliavin \cite{GM}, as refined in \cite{DH1}, allows the
Segal--Bargmann transform for the compact group $K$ to be viewed as a special
case of the Segal--Bargmann transform for an infinite-dimensional Euclidean
space. In Section \ref{inf.sec}, I explain how the results of Section
\ref{mult.sec} can be derived, at least formally, from the
infinite-dimensional perspective.

I am grateful to Bruce Driver and Laurent Saloff-Coste for very helpful
discussions. I also thank the referee for corrections which have improved the
quality of the paper.

\section{Berezin--Toeplitz and Schr\"{o}dinger quantization for $T^{\ast}(K)
$\label{schr.sec}}

The Hilbert space we will consider will ultimately be identified with the
space of square-integrable holomorphic functions on $K_{\mathbb{C}}$ with
respect to a $K$-invariant heat kernel measure $\nu_{t}(g)~dg.$ This space is
denoted $\mathcal{H}L^{2}(K_{\mathbb{C}},\nu_{t}).$ As I will now explain,
however, this space can also be obtained by applying the method of geometric
quantization with half-forms, thus connecting with much of the literature on
Berezin--Toeplitz quantization. (See, for example, the work \cite{Ra} of J.
Rawnsley, who interprets the work of Berezin \cite{Be} in terms of geometric quantization.)

Let $K$ be a connected compact Lie group, assumed to be semisimple. We choose
once and for all a bi-invariant Riemannian metric on $K.$ This is equivalent
to choosing an Ad-invariant inner product on the Lie algebra $\mathfrak{k}$ of
$K.$ There is then a natural \textquotedblleft adapted complex
structure\textquotedblright\ on $T^{\ast}(K),$ described independently and in
slightly different language by \cite{GStenz1,GStenz2} and \cite{LS,Sz1}. This
complex structure fits together with the canonical symplectic structure on
$T^{\ast}(K)$ to form a K\"{a}hler structure. Furthermore, $T^{\ast}(K)$ with
its adapted complex structure is biholomorphic in a natural way to the
complexification $K_{\mathbb{C}}$ of $K $ \cite[Sect. 3]{phasebounds}. Here
$K_{\mathbb{C}}$ is the unique connected complex Lie group that has Lie
algebra $\mathfrak{k}_{\mathbb{C}}:=\mathfrak{k}+i\mathfrak{k}$ and that
contains $K$ as a maximal compact subgroup; for example, if $K=SU(n)$ then
$K_{\mathbb{C}}=SL(n,\mathbb{C}).$

We now apply the method of geometric quantization (with half-forms) with
respect to the adapted complex structure. This amounts to constructing a
certain holomorphic line bundle over $T^{\ast}(K)$ and giving a certain recipe
for computing the norm of such a section. The quantum Hilbert space is then
the space of holomorphic sections of finite norm. In the case at hand, this
bundle is holomorphically trivial. Upon choosing a natural trivialization, the
quantum Hilbert space becomes the space of holomorphic functions that are
square integrable with respect to a certain measure. If we identify $T^{\ast
}(K)$ with $K_{\mathbb{C}}$ then this measure turns out to coincide, up to an
irrelevant constant, with the $K$-invariant heat kernel measure $\nu
_{t}(g)~dg$ considered in \cite{H1}. (Details on the definition of $\nu_{t}$
will be given in Section \ref{mult.sec}.) Here $dg$ is a Haar measure on
$K_{\mathbb{C}}$ and $t$ is a positive parameter that plays the role of
Planck's constant. See \cite{geoquant} for the details of this calculation.
Our quantum Hilbert space is thus identified with the space of holomorphic
functions on $K_{\mathbb{C}}$ that are square integrable with respect to
$\nu_{t}.$ We denote this space $\mathcal{H}L^{2}(K_{\mathbb{C}},\nu_{t})$ and
refer to it as the Segal--Bargmann space.

Now that we have the Segal--Bargmann space $\mathcal{H}L^{2}(K_{\mathbb{C}%
},\nu_{t}),$ we consider Toeplitz operators on it. Let $P_{t}$ denote the
orthogonal projection from $L^{2}(K_{\mathbb{C}},\nu_{t})$ onto the
holomorphic subspace. If $\phi$ is a bounded measurable function, then we
define the Toeplitz operator $T_{\phi}$ with symbol $\phi$, as an operator
from $\mathcal{H}L^{2}(K_{\mathbb{C}},\nu_{t})$ to itself, by%
\[
T_{\phi}(F)=P_{t}(\phi F).
\]
That is, $T_{\phi}$ consists of multiplication by $\phi$ followed by
projection back into the holomorphic subspace. If $\phi$ is an unbounded
measurable function, $T_{\phi}$ may still be defined by the same formula, but
restricted to the domain of those $F$'s for which $\phi F$ is in
$L^{2}(K_{\mathbb{C}},\nu_{t}).$ The operator $T_{\phi}$ will typically be
unbounded. In the cases we will consider in this paper, $T_{\phi}$ will always
be a densely defined operator on $\mathcal{H}L^{2}(K_{\mathbb{C}},\nu_{t}).$

As an alternative to the Segal--Bargmann space, one has the
Schr\"{o}dinger-type Hilbert space, $L^{2}(K).$ (If no other measure is
specified, $L^{2}(K)$ is understood to be with respect to the Riemannian
volume measure $dx$, which is a Haar measure.) Use of $L^{2}(K)$ as the
Hilbert space leads to a natural way of quantizing certain functions. For
example, suppose $\phi$ is a function on $K_{\mathbb{C}}\cong T^{\ast}(K)$
that is constant along each cotangent space. Then $\phi$ is of the form
$\phi=V\circ\pi,$ where $\pi$ is the projection from $T^{\ast}(K)$ to $K.$ It
is natural to quantize such a function as multiplication by $V$ acting on
$L^{2}(K).$ Similarly, functions on $T^{\ast}(K)$ that are polynomials of
degree at most $N$ on each fiber get quantized as differential operators of
degree at most $N$ acting as (unbounded) operators on $L^{2}(K).$

In \cite{H1}, I introduced a generalized Segal--Bargmann transform for $K,$
which is a unitary map $C_{t}$ of $L^{2}(K)$ onto $\mathcal{H}L^{2}%
(K_{\mathbb{C}},\nu_{t}).$ This operator is defined by applying the heat
operator $e^{t\Delta/2}$ to a function $f$ in $L^{2}(K)$ and then analytically
continuing the resulting function to $K_{\mathbb{C}}.$ In \cite{geoquant}, I
show that $C_{t}$ coincides (up to a constant) with the \textquotedblleft
pairing map\textquotedblright\ of geometric quantization.

Since we have a natural unitary map between $L^{2}(K)$ and $\mathcal{H}%
L^{2}(K_{\mathbb{C}},\nu_{t}),$ it is natural to compare the quantization
procedures associated to these two Hilbert spaces. The main goal of the
present paper is to demonstrate how multiplication operators and differential
operators on $L^{2}(K),$ when conjugated by the Segal--Bargmann transform, can
be expressed as Toeplitz operators on $\mathcal{H}L^{2}(K_{\mathbb{C}},\nu
_{t}).$

\section{Multiplication operators\label{mult.sec}}

In this section, we will consider multiplication by a function $V:K\rightarrow
\mathbb{C}$ as an operator on $L^{2}(K)$ and then conjugate this operator by
the Segal--Bargmann transform. If $V$ is sufficiently regular, we will show
that $C_{t}M_{V}C_{t}^{-1}$ can be expressed as a Toeplitz operator with
symbol $\phi_{V}$ and give a formula for $\phi_{V}$ in terms of $V.$ This
formula involves a certain subelliptic heat kernel on $K_{\mathbb{C}}.$

It is helpful to consider first the $\mathbb{R}^{n}$ case, with a
Segal--Bargmann transform $C_{t}$ mapping from $L^{2}(\mathbb{R}^{n})$ onto
$\mathcal{H}L^{2}(\mathbb{C}^{n},\nu_{t}),$ where
\[
\nu_{t}(x+iy)=(\pi t)^{-d/2}e^{-\left\vert y\right\vert ^{2}/t}.
\]
Here $C_{t}f$ is the analytic continuation (from $\mathbb{R}^{n}$ to
$\mathbb{C}^{n}$) of $e^{t\Delta/2}f.$ See \cite{mexnotes} or \cite[Sect.
3]{newform} for a comparison of the normalization conventions I am using here
to those used in \cite{Ba1,Se3}.

It is well known (e.g., Propositions 2.96 and 2.97 of \cite{Fo}) that for
reasonable symbols $\phi,$ the operator $C_{t}^{-1}T_{\phi}C_{t}$ coincides
with the Weyl quantization of the function $\hat{\phi},$ where $\hat{\phi
}=e^{t\Delta/4}\phi.$ If $\phi(x+iy)$ depends only on $x,$ then the same is
true of $\hat{\phi}(x+iy).$ In that case, the Weyl quantization of $\hat{\phi
}$ is simply the operation of multiplication by $\hat{\phi}(x),$ acting on
$L^{2}(\mathbb{R}^{n}).$ Thus, given a multiplication operator $M_{V}$ on
$L^{2}(\mathbb{R}^{n}),$ if there exists a function $\tilde{V}$ such that
$e^{t\Delta/4}\tilde{V}=V,$ then we have
\begin{equation}
C_{t}M_{V}C_{t}^{-1}=T_{\tilde{V}}.\label{rn.form}%
\end{equation}
On the right-hand side of (\ref{rn.form}), we abuse notation slightly and
allow $\tilde{V}$ to stand for the function on $\mathbb{C}^{n}$ given by
$x+iy\longmapsto\tilde{V}(x).$ Of course, in order for such a $\tilde{V}$ to
exist, $V$ itself must be extremely regular. (See, for example, \cite{Hi} or
\cite{range} for some discussion of how regular $V$ must be.)

We now proceed to the case of a connected compact Lie group $K$. In the
interests of notational simplicity, we assume $K$ is semisimple. The results
in the torus case are essentially the same as in the $\mathbb{R}^{n}$ case,
whereas the semisimple case involves a subelliptic heat kernel that does not
show up in the $\mathbb{R}^{n}$ case.

We fix an Ad-invariant inner product on $\mathfrak{k}$, which determines a
bi-invariant Riemannian metric on $K$. We let $dx$ denote the Riemannian
volume measure on $K$, which is a Haar measure. We let $\Delta_{K}$ denote the
Laplacian with respect to this metric, take to be a \textit{negative}
operator. This operator can be computed as $\Delta_{K}=\sum X_{k}^{2},$ where
the $X_{k}$'s form an orthonormal basis for $\mathfrak{k}$ and are viewed as
left-invariant differential operators. We let $\rho_{t}$ be the fundamental
solution at the identity of the heat equation $\partial\rho_{t}/\partial
t=\frac{1}{2}\Delta_{K}\rho_{t}.$ For each fixed $t>0,$ the function $\rho
_{t}$ admits an analytic continuation (also denoted $\rho_{t} $) to
$K_{\mathbb{C}}.$ Let $\mathcal{H}(K_{\mathbb{C}})$ denote the space of
holomorphic functions on $K_{\mathbb{C}}$ and let $C_{t}$ be the map from
$L^{2}(K)$ (with respect $dx$) into $\mathcal{H}(K_{\mathbb{C}})$ given by%
\begin{equation}
C_{t}f(g)=\int_{K}\rho_{t}(gx^{-1})f(x)~dx,\quad g\in K_{\mathbb{C}%
}.\label{ct.form}%
\end{equation}

If $\{X_{k}\}_{k=1}^{\dim K}$ is an orthonormal basis for $\mathfrak{k},$ then
$\left\{  X_{k},JX_{k}\right\}  _{k=1}^{\dim K}$ forms a basis of
$\mathfrak{k}_{\mathbb{C}}.$ We now regard each $X_{k}$ and each $JX_{k}$ as a
left-invariant differential operator on $K_{\mathbb{C}}.$ The function
$\nu_{t}$ is the solution (in $L^{2}(K_{\mathbb{C}},dg)$) to the heat equation%
\[
\frac{\partial\nu}{\partial t}=\frac{1}{4}\sum_{k=1}^{\dim K}\left(  X_{k}%
^{2}+(JX_{k})^{2}\right)  \nu_{t}
\]
subject to the initial condition%
\[
\lim_{t\rightarrow0^{+}}\int_{K_{\mathbb{C}}}f(g)\nu_{t}(g)~dg=\int
_{K}f(x)~dx.
\]
Here $dg$ is a fixed Haar measure on $K_{\mathbb{C}}.$

The function $\nu_{t}$ is the heat kernel at the identity coset for the
symmetric space $K_{\mathbb{C}}/K,$ viewed as a right-$K$-invariant function
on $K_{\mathbb{C}}.$ We let $L^{2}(K_{\mathbb{C}},\nu_{t})$ denote the $L^{2}$
space with respect to the measure $\nu_{t}(g)~dg,$ and we let $\mathcal{H}%
L^{2}(K_{\mathbb{C}},\nu_{t})$ denote the holomorphic subspace thereof.
According to Theorem 2 of \cite{H1}, $C_{t}$ is a unitary map of $L^{2}(K)$
onto $\mathcal{H}L^{2}(K_{\mathbb{C}},\nu_{t}).$ We let $P_{t}$ denote the
orthogonal projection of $L^{2}(K_{\mathbb{C}},\nu_{t})$ onto the holomorphic
subspace. For any bounded measurable function $\phi$ on $K_{\mathbb{C}},$ we
let $T_{\phi}$ denote the Toeplitz operator on $\mathcal{H}L^{2}%
(K_{\mathbb{C}},\nu_{t})$ given by $T_{\phi}(F)=P_{t}(\phi F).$ If $F_{1}$ and
$F_{2}$ belong to $\mathcal{H}L^{2}(K_{\mathbb{C}},\nu_{t})$ then
\begin{equation}
\left\langle F_{1},T_{\phi}F_{2}\right\rangle _{\mathcal{H}L^{2}%
(K_{\mathbb{C}},\nu_{t})}=\int_{K_{\mathbb{C}}}F_{1}(g)\phi(g)F_{2}(g)\nu
_{t}(g)~dg\label{matrix.entries}%
\end{equation}
because $P_{t}$ is self-adjoint and $P_{t}F_{1}=F_{1}.$

In \cite{DH1} (see also \cite{newform}), B. Driver and I consider a family
$A_{s,t}$ of operators on $K_{\mathbb{C}}$ parameterized by two positive
numbers $s$ and $t$ with $s\geq t/2,$%
\[
A_{s,t}=\left(  s-\frac{t}{2}\right)  \sum_{k=1}^{\dim K}X_{k}^{2}+\frac{t}%
{2}\sum_{k=1}^{\dim K}(JX_{k}{})^{2},
\]
where $J$ is the \textquotedblleft multiplication by $i$\textquotedblright%
\ map on $\mathfrak{k}_{\mathbb{C}}.$ If $s>t/2,$ then $A_{s,t}$ is an
elliptic operator. We now consider the heat equation $\partial u/\partial
r=\frac{1}{2}A_{s,t}u$ on $K_{\mathbb{C}}$ and we let $\mu_{s,t}$ denote the
fundamental solution of this equation at the identity, evaluated at $r=1.$
This is equivalent to saying that%
\[
\mu_{s,t}=e^{A_{s,t}/2}(\delta_{e}),
\]
where $\delta_{e}$ is a Dirac delta-function at the identity.

Since we are assuming that $K$ is semisimple, the \textquotedblleft
borderline\textquotedblright\ operator $A_{t/2,t}$ satisfies H\"{o}rmander's
condition and is therefore hypoelliptic. In the semisimple case, the heat
kernel $\mu_{s,t}$ is still a well-defined, smooth, strictly positive function
on $K_{\mathbb{C}}$ when $s=t/2.$ The subelliptic heat kernel%
\begin{equation}
\mu_{t/2,t}=\exp\left\{  \frac{t}{4}\sum_{k=1}^{\dim K}(JX_{k})^{2}\right\}
(\delta_{e})\label{mut.form}%
\end{equation}
plays an essential role in all the main results of this paper.

Meanwhile, the Casimir operator $\sum((JX_{k})^{2}-X_{k}^{2})$ commutes with
each $X_{j}$ and each $JX_{j}.$ It follows that $\sum X_{k}^{2}$ commutes with
$\sum(JX_{k})^{2}$. We then have (at least formally)%
\[
e^{A_{s,t}/2}=e^{A_{s-r,t}/2}e^{r\Delta_{K}/2}.
\]
It is then not hard to show (cf. \cite[Sect. 8]{H1}) that for $s-r\geq t/2$ we
have%
\begin{equation}
\mu_{s,t}(g)=\int_{K}\mu_{s-r,t}(gx^{-1})\rho_{r}(x)~dx,\quad g\in
K_{\mathbb{C}}.\label{mustr}%
\end{equation}

For any $s\geq t/2$ we have%
\begin{equation}
\nu_{t}(g)=\int_{K}\mu_{s,t}(gx^{-1})~dx.\label{mustnut}%
\end{equation}
Furthermore,%
\begin{equation}
\lim_{s\rightarrow\infty}\mu_{s,t}(g)=\mathrm{Vol}(K)\nu_{t}%
(g),\label{must.inf}%
\end{equation}
where $\mathrm{Vol}(K)$ is the Riemannian volume of $K.$

We are now ready to state the main result of this section.

\begin{theorem}
\label{ss.thm}Suppose that $V$ is a function on $K$ and that there exists a
bounded measurable function $\tilde{V}$ on $K$ such that $V=e^{t\Delta
/4}\tilde{V}.$ Let $\phi_{V}$ be the function on $K_{\mathbb{C}}$ defined by
\[
\phi_{V}(g)=\frac{\int_{K}\mu_{t/2,t}(gx^{-1})\tilde{V}(x)~dx}{\nu_{t}(g)}.
\]
Then $\phi_{V}$ is a bounded function and
\[
C_{t}M_{V}C_{t}^{-1}=T_{\phi_{V}}.
\]

\end{theorem}

In the commutative case, the formula would be $\phi_{V}(xe^{iY})=\tilde
{V}(x),$ essentially the same as what we have in the $\mathbb{R}^{n}$ case.

\begin{proof}
Applying (\ref{mustnut}) with $s=t/2,$ we see that%
\[
\left\vert \phi_{V}(g)\right\vert \leq\sup|\tilde{V}|,
\]
establishing the boundedness of $\phi_{V}.$

Since $C_{t}$ (as defined in (\ref{ct.form})) is an integral operator, its
adjoint is easily computed as%
\begin{equation}
(C_{t}^{\ast}\Phi)(x)=\lim_{n\rightarrow\infty}\int_{E_{n}}\overline{\rho
_{t}(gx^{-1})}\Phi(g)~dg,\label{ct.adj}%
\end{equation}
for any $\Phi\in L^{2}(K_{\mathbb{C}},\nu_{t}).$ Here $E_{n}$ is any
increasing sequence of compact, $K$-invariant subsets of $K_{\mathbb{C}}$
whose union is $K_{\mathbb{C}}$ and the limit is in the norm topology of
$L^{2}(K).$ (See \cite[Sect. 8]{H1}.)

For any fixed $s,$ we consider the heat kernel measure $\rho_{s}(x)~dx$ and
the resulting $L^{2}$ space, which we denote $L^{2}(K,\rho_{s}).$ We assume
for the moment that $s>t/2$ (we will eventually let $s$ tend to $t/2$) and we
consider the transform denoted $B_{s,t}$ in \cite{DH1,newform}. This is the
map from $L^{2}(K,\rho_{s})$ into $\mathcal{H}(K_{\mathbb{C}})$ given by%
\begin{align}
(B_{s,t}f)(g)  & =\int_{K}\rho_{t}(gx^{-1})f(x)~dx\nonumber\\
& =\int_{K}\frac{\rho_{t}(gx^{-1})}{\rho_{s}(x)}f(x)~\rho_{s}%
(x)~dx.\label{bt.form}%
\end{align}
Note that the formula for $B_{s,t}$ is the same as the formula for $C_{t}$ and
is independent of $s$; only the inner product on the domain space depends on
$s.$

According to \cite[Thm. 5.3]{DH1} or \cite[Thm. 1.2]{newform}, $B_{s,t}$ is an
isometric map of $L^{2}(K,\rho_{s})$ into $L^{2}(K_{\mathbb{C}})$ with respect
to the measure $\mu_{s,t}(g)~dg,$ whose image is precisely the holomorphic
subspace $\mathcal{H}L^{2}(K_{\mathbb{C}},\mu_{s,t}).$ Since $B_{s,t}$ is
isometric, its adjoint is a one-sided inverse, where the adjoint is readily
computed from (\ref{bt.form}). Given $f\in L^{2}(K,\rho_{s}),$ if we let
$F=B_{s,t}f$ then we have an inversion formula given by%
\begin{equation}
f(x)=\lim_{n\rightarrow\infty}\int_{E_{n}}F(g)\frac{\overline{\rho_{t}%
(gx^{-1})}}{\rho_{s}(x)}\mu_{s/2,t}(g)~dg.\label{inv}%
\end{equation}
Since $\rho_{s}(x)$ is independent of $g,$ we may pull this factor outside the
integral in (\ref{inv}) and then multiply both sides by $\rho_{s}$ to obtain%
\begin{align}
\rho_{s}(x)f(x)  & =\lim_{n\rightarrow\infty}\int_{E_{n}}F(g)\overline
{\rho_{t}(gx^{-1})}\mu_{s,t}(g)~dg\nonumber\\
& =\lim_{n\rightarrow\infty}\int_{E_{n}}\frac{\mu_{s,t}(g)}{\nu_{t}%
(g)}F(g)\overline{\rho_{t}(gx^{-1})}\nu_{t}(g)~dg.\label{rhosf}%
\end{align}

Now, the proof of the \textquotedblleft averaging lemma\textquotedblright\ in
\cite{H1} applies to $\mu_{s,t}$ for $s>t/2$ and yields that for each fixed
$s,t$ with $s>t/2$ we have positive constants $c_{1}$ and $c_{2}$ such that%
\[
c_{1}\nu_{t}(g)\leq\mu_{s,t}(g)\leq c_{2}\nu_{t}(g)
\]
for all $g\in K_{\mathbb{C}}.$ (This result follows fairly easily from
(\ref{mustr})\ and (\ref{mustnut}).) It follows that the function $\Phi$ on
$K_{\mathbb{C}}$ given by%
\[
\Phi(g)=\frac{\mu_{s,t}(g)}{\nu_{t}(g)}F(g)
\]
belongs to $L^{2}(K_{\mathbb{C}},\nu_{t}).$ Comparing (\ref{rhosf}) to
(\ref{ct.adj}) we obtain%
\[
\rho_{s}f=C_{t}^{\ast}\Phi.
\]

Now, since $C_{t}$ is isometric and its image is precisely the holomorphic
subspace $\mathcal{H}L^{2}(K_{\mathbb{C}},\nu_{t}),$ the adjoint map may be
computed as%
\[
C_{t}^{\ast}=C_{t}^{-1}P_{t},
\]
where $P_{t}$ is the orthogonal projection of $L^{2}(K_{\mathbb{C}},\nu_{t}) $
onto $\mathcal{H}L^{2}(K_{\mathbb{C}},\nu_{t}).$ Here $C_{t}^{-1}$ denotes the
inverse of $\ C_{t}$ as a map of $L^{2}(K)$ onto $\mathcal{H}L^{2}%
(K_{\mathbb{C}},\nu_{t}).$ We conclude, then, that%
\begin{equation}
\rho_{s}f=C_{t}^{-1}P_{t}\left(  \frac{\mu_{s,t}}{\nu_{t}}F\right)
.\label{equality}%
\end{equation}

Now, the density $\rho_{s}$ is bounded and bounded away from zero, so if $f$
belongs to $L^{2}(K,\rho_{s})$ then it also belongs to $L^{2}(K).$
Furthermore, the formula for $B_{s,t}f$ is the same as for $C_{t}f.$ Thus,
$F,$ which was defined to be $B_{s,t}f$, coincides with $C_{t}f$ and
(\ref{equality}) becomes $\rho_{s}f=C_{t}^{-1}T_{\phi}C_{t}f,$ where $\phi
=\mu_{s,t}/\nu_{t}.$ Thus,%
\begin{equation}
C_{t}M_{\rho_{s}}C_{t}^{-1}=T_{\phi},\quad\phi=\frac{\mu_{s,t}}{\nu_{t}%
}.\label{rhos.form}%
\end{equation}
Since the transform $C_{t}$ commutes with left- and right-translations by
elements of $K$ (since $\Delta_{K}$ is bi-invariant), it is easy to see that%
\begin{equation}
C_{t}M_{R_{x}\rho_{s}}C_{t}^{-1}=T_{\phi_{x}},\quad\phi_{x}=\frac{R_{x}%
\mu_{s,t}}{\nu_{t}}\label{ry}%
\end{equation}
where for any function $f$ on $K$ or $K_{\mathbb{C}},$ we set $(R_{y}%
f)(g)=f(gy^{-1}).$ (Recall that $\nu_{t}$ is invariant under the right action
of $K.$)

Recall now that $V$ is a function on $K$ of the form $V=e^{t\Delta/4}\tilde
{V},$ where $\tilde{V}$ is a bounded measurable function. For $s>t/2,$ let us
integrate (\ref{ry}) against $\tilde{V}.$ (On the right-hand side, write
things in terms of the matrix entries as in (\ref{matrix.entries}) and use
Fubini.) Note that $e^{s\Delta/2}\tilde{V}$ may be computed as $\int
_{K}\left(  R_{x}\rho_{s}\right)  f(x)~dx.$ We obtain, then,%
\begin{equation}
C_{t}M_{e^{s\Delta_{K}/2}\tilde{V}}C_{t}^{-1}=T_{\phi_{s,V}}\label{s.form}%
\end{equation}
where%
\begin{equation}
\phi_{s,V}(g)=\frac{\int_{K}\mu_{s,t}(gx^{-1})\tilde{V}(x)~dx}{\nu_{t}%
(g)}\label{phis.form}%
\end{equation}

Now, as $s$ decreases to $t/2,$ $e^{s\Delta_{K}/2}\tilde{V}$ converges
uniformly to $e^{t\Delta_{K}/4}\tilde{V}=V.$ Meanwhile, from (\ref{mustr})
with $s=t/2$ we see that $\mu_{s,t}$ tends pointwise to $\mu_{t/2,t}$ as $s $
decreases to $t/2.$ Furthermore, (\ref{mustnut}) tells us that $\left\vert
\phi_{s,V}(g)\right\vert $ is bounded by $\sup|\tilde{V}|,$ independently of
$s.$ It then follows from Dominated Convergence (and (\ref{matrix.entries}))
that $T_{\phi_{s,V}}$ tends weakly to $T_{\phi_{V}}$ as $s\rightarrow t/2.$
Thus, letting $s$ decrease to $t/2$ in (\ref{s.form}) gives the desired result.
\end{proof}

\section{Differential operators\label{diff.sec}}

In this Section, we will consider differential operators acting as unbounded
operators on $L^{2}(K).$ We \textit{do} not assume that the operators are
right- or left-invariant, but we do assume that the coefficients (say, when
expanded in terms of left-invariant vector fields) are very regular. By
differentiating the results of Section \ref{mult.sec} and then integrating by
parts, we will see that a differential operator, when conjugated by the
Segal--Bargmann transform, can be expressed as a Toeplitz operator.

Now, in the previous section, under sufficiently stringent assumptions on $V,
$ we were able to express $C_{t}M_{V}C_{t}^{-1}$ as a Toeplitz operator with a
bounded symbol. Differential operators, however, are unbounded and the
corresponding Toeplitz symbols are necessarily unbounded as well. In general,
we will \textit{not} obtain equality of domains between a differential
operator $\alpha$ (conjugated by $C_{t}$) and the associated Toeplitz
operator. (Nevertheless, see \cite[Rem. 19]{HL} for examples where equality of
domains does occur.) Rather, we will content ourselves by establishing
equality of the operators on the space of finite linear combinations of matrix entries.

By the results of Section \ref{mult.sec}, we have%
\[
Vf=C_{t}^{-1}P_{t}M_{\phi_{V}}C_{t}f
\]
for sufficiently nice $V$ and $f\in L^{2}(K).$ Letting $F=C_{t}f$ and
recalling that $C_{t}^{\ast}=C_{t}^{-1}P_{t},$ we have%
\begin{equation}
V(x)f(x)=\lim_{n\rightarrow\infty}\int_{E_{n}}F(g)\overline{\rho_{t}(gx^{-1}%
)}\int_{K}\mu_{t/2,t}(gy^{-1})\tilde{V}(y)~dy~dg.\label{v.inv}%
\end{equation}
(Compare (\ref{ct.adj}).)

Now let $A$ be a left-invariant differential operator on $K,$ a linear
combination of products of left-invariant vector fields. Those left-invariant
vector fields can be extended to left-invariant vector fields on
$K_{\mathbb{C}},$ and so $A$ may be regarded as a left-invariant operator on
$K_{\mathbb{C}}.$ Since $C_{t}$ commutes with left- and right-translations by
elements of $K,$ $C_{t}Af=AC_{t}f.$ We may therefore replace $f$ by $Af$ and
$F$ by $AF$ in (\ref{v.inv}), assuming $f$ is in the domain of $A.$

For a left-invariant vector field $X,$ let $X_{\mathbb{C}}$ be the holomorphic
vector field given by%
\[
X_{\mathbb{C}}=\frac{1}{2}(X-iJX).
\]
Then $X_{\mathbb{C}}F=XF$ if $F$ is holomorphic and $X_{\mathbb{C}}F=0$ if $F
$ is antiholomorphic. If $A$ is written as a sum of products of left-invariant
vector fields, we may produce a holomorphic differential operator
$A_{\mathbb{C}}$ by replacing each vector field $X$ by $X_{\mathbb{C}}.$ Then
$A_{\mathbb{C}}F=F$ for all holomorphic functions $F.$

Let us now replace $f$ by $Af$ and $F$ by $AF$ in (\ref{v.inv}), and then
change $AF$ to $A_{\mathbb{C}}F.$ We obtain, then,%
\begin{equation}
V(x)Af(x)=\lim_{n\rightarrow\infty}\int_{E_{n}}\left(  A_{\mathbb{C}%
}F(g)\right)  \overline{\rho_{t}(gx^{-1})}\int_{K}\mu_{t/2,t}(gy^{-1}%
)\tilde{V}(y)~dy~dg.\label{va.inv}%
\end{equation}

We now want to integrate by parts on the right-hand side of (\ref{va.inv}).
Let $B\longmapsto B^{tr}$ be the linear (\textit{not} conjugate-linear) map on
left-invariant operators satisfying%
\[
\left(  X_{1}X_{2}\cdots X_{N}\right)  ^{tr}=(-1)^{N}X_{N}\cdots X_{2}X_{1}.
\]
If $F$ is nice enough, the limit in (\ref{va.inv}) will become simply
integration over $K_{\mathbb{C}}.$ If we then integrate by parts repeatedly,
assuming boundary terms may be neglected, we get%
\begin{align}
V(x)Af(x)  & =\int_{K_{\mathbb{C}}}F(g)\overline{\rho_{t}(gx^{-1})}\left[
A_{\mathbb{C}}^{tr}\int_{K}\mu_{t/2,t}(gy^{-1})\tilde{V}(y)~dy\right]
~dg\nonumber\\
& =\int_{K_{\mathbb{C}}}F(g)\overline{\rho_{t}(gx^{-1})}\phi_{V,A}(g)\nu
_{t}(g)~dg,\label{va.inv2}%
\end{align}
where%
\begin{equation}
\phi_{V,A}=\frac{A_{\mathbb{C}}^{tr}\int_{K}\mu_{t/2,t}(gx^{-1})\tilde
{V}(x)~dx}{\nu_{t}(g)}.\label{phiva.form}%
\end{equation}
Note that since $\overline{\rho_{t}(gx^{-1})}$ is antiholomorphic as a
function of $g,$ the holomorphic operator $A_{\mathbb{C}}^{tr}$ does not
\textquotedblleft see\textquotedblright\ this factor.

Assume that the boundary terms in the integration by parts may indeed be
neglected. Assume also that $F\phi_{V,A}$ is in $L^{2}(K_{\mathbb{C}},\nu
_{t}).$ Then (\ref{va.inv2}) tells us that%
\begin{align*}
M_{V}Af  & =C_{t}^{\ast}M_{\phi_{V,A}}C_{t}f\\
& =C_{t}^{-1}T_{\phi_{V,A}}C_{t}f.
\end{align*}
Thus, on some as yet undetermined domain in $\mathcal{H}L^{2}(K_{\mathbb{C}%
},\nu_{t}),$ we have%
\[
C_{t}M_{V}AC_{t}^{-1}=T_{\phi_{V,A}}.
\]

If $A=I,$ then this simply reproduces the results of Section \ref{mult.sec}.
On the other hand, if $V$ is the constant function $\mathbf{1},$ then by
(\ref{mustnut})%
\[
\phi_{\mathbf{1},A}=\frac{A_{\mathbb{C}}^{tr}\nu_{t}}{\nu_{t}},
\]
reproducing a result of \cite{HL}. Because there is a fairly simple explicit
expression for $\nu_{t},$ it is possible to compute $\phi_{\mathbf{1},A}$
quite explicitly in some cases. For example, it easy to see that if $A$ is a
power of the Laplacian, then $\phi_{\mathbf{1},A}(xe^{iY})$ is a polynomial in
$\left\vert Y\right\vert ^{2}.$ See \cite[Sect. 3]{HL} for more information.

\begin{theorem}
\label{diff.thm}Suppose $f$ is a finite linear combination of matrix entries
and let $F=C_{t}f.$ Suppose $V$ is of the form $V=e^{t\Delta_{K}/4}\tilde{V} $
for some bounded measurable function $\tilde{V}$ on $K.$ Then $\phi_{V,A}F $
belongs to $L^{2}(K_{\mathbb{C}},\nu_{t})$ and%
\[
C_{t}^{-1}P_{t}(\phi_{V,A}F)=M_{V}Af,
\]
where $\phi_{V,A}$ is given in (\ref{phiva.form}).
\end{theorem}

\begin{proof}
The main issue in the proof of Theorem \ref{diff.thm} is to obtain reasonable
estimates on the subelliptic heat kernel $\mu_{t/2,t}$ and its derivatives.
Such estimates can be obtained by using an appropriate parabolic Harnack
inequality, such as Theorem V.3.1 in \cite{VSC}. This inequality bounds
derivatives of the heat kernel (or any positive solution of the heat equation)
at fixed point and some time $t$ by a constant times the heat kernel itself at
the same point and some slightly later time. In the group case, the
homogeneity of the problem means that the constant can be taken to be
independent of the point. Thus, for any $\tau>t>0$ and any left-invariant
differential operator $\alpha,$ we have a constant $c$ (depending on $t,$
$\tau,$ and $\alpha$ but not on $g$) such that%
\[
\alpha\mu_{t/2,t}(g)\leq c\mu_{\tau/2,\tau}(g)
\]
for all $g\in K_{\mathbb{C}}.$

We now make use of the pointwise bounds for $\mu_{\tau/2,\tau},$ which are
expressed in terms of a sub-Riemannian distance function on $K_{\mathbb{C}}.$
The distance is the infimum of lengths of paths joining two points, where we
allow only paths whose tangent vectors at each point lie in the span of the
left-invariant vector fields $\{JX_{k}\}_{k=1}^{\dim K}$. (Any two points can
be joined by such a path.) The length of an allowed path is computed by
identifying $J\mathfrak{k}$ with $\mathfrak{k}$ and using the fixed
Ad-invariant inner product on $\mathfrak{k}.$

Let us think about this distance function in terms of the polar decomposition
for $K_{\mathbb{C}},$ which expresses each $g\in K_{\mathbb{C}} $ uniquely as
$g=xe^{iY},$ with $x\in K$ and $Y\in\mathfrak{k}.$ The length of any allowed
path in $K_{\mathbb{C}}$ is the equal to the length of its projection into
$K_{\mathbb{C}}/K,$ where we use on $K_{\mathbb{C}}/K$ an obvious
left-$K_{\mathbb{C}}$-invariant Riemannian metric. What this means is that
\[
d(e,xe^{iY})\geq d(eK,xe^{iY}K)=|Y|.
\]
Meanwhile, the distance from $e$ to $x\in K$ is bounded, from which it follows
that $d(e,xe^{iY})\leq|Y|+C.$

Meanwhile, according to Theorem VIII.4.3 of \cite{VSC}, for all $\varepsilon
>0$ there exists $C_{\varepsilon}$
\[
\mu_{t/2,t}(g)\leq C_{\varepsilon}t^{-n/2}e^{-d(e,g)^{2}/(t+\varepsilon)},
\]
where $n$ is the \textquotedblleft local dimension.\textquotedblright\ (The
absence of a 4 in the denominator in the exponent is due to our normalization
of the subelliptic Laplacian; see (\ref{mut.form}).) On the other hand, there
is an explicit formula for $\nu_{t}(xe^{iY})$ (due to R. Gangolli \cite{Ga})
and it is a Gaussian in $Y,$ multiplied by an exponentially decaying factor.
So $1/\nu_{t}$ is bounded by $e^{|Y|^{2}/t}$ times a factor that grows no more
than exponentially.

So, $\mu_{t/2,t}(xe^{iY})$ is bounded by a constant times a Gaussian in $Y,$
where the constant in the exponent of the Gaussian is arbitrarily close to
what one has in the Euclidean case. By the parabolic Harnack inequality, the
same is true of any left-invariant derivatives of $\mu_{t/2,t}.$ From this and
the bounds on $\nu_{t},$ we see that $\phi_{V,A}(xe^{iY})$ is bounded by a
constant times a Gaussian in $\left\vert Y\right\vert ,$ where the constant in
the exponent in the Gaussian can be made as close to zero as we like.

Now, $\rho_{t}(xe^{iY})$ can be bounded using the calculations in
\cite{phasebounds} (compare also \cite{Tha}), by $e^{|Y|^{2}/2t}$ times an
exponentially decaying factor. All of these estimates together show that the
integral in (\ref{va.inv2}) is absolutely convergent. (The function $F$ has at
most exponential growth, $\rho_{t}$ grows like $e^{|Y|^{2}/2t},$ $\phi_{V,A}$
grows at most like $e^{\varepsilon|Y|^{2}}$ and $\nu_{t}$ decays like
$e^{-|Y|^{2}/t}.$) Similar remarks apply to the integral on the right-hand
side of (\ref{va.inv}), which means that the limit can be replaced by an
integral over all of $K_{\mathbb{C}}.$ Using the arguments in Section 4 of
\cite{HL}, there is no problem in justifying the integration by parts to
establish the correctness of (\ref{va.inv2}). Since also $\phi_{V,A}F\in
L^{2}(K_{\mathbb{C}},\nu_{t}),$ (\ref{va.inv2}) amounts to saying that
$C_{t}^{-1}P_{t}(\phi_{V,A}F)=VAf$
\end{proof}

\section{The infinite-dimensional perspective\label{inf.sec}}

In this section, we will look at the results of Section \ref{mult.sec} from an
infinite-dimensional point of view. (Presumably the point of view could be
extended to the results of Section \ref{diff.sec}, but I will not consider
that problem here.) We will derive (nonrigorously) a formula for the Toeplitz
symbol $\phi_{V}$ in terms of Gaussian measures and the It\^{o} map, and then
verify (rigorously) that this infinite-dimensional prediction indeed
reproduces the results of Theorem \ref{ss.thm}. See (\ref{phiv.inf}) and
(\ref{lim}). In particular, the measure $\mu_{t/2,t}(g)~dg$ can be seen to
arise from a certain application of the It\^{o} map.

\subsection{The work of Gross and Malliavin}

The motivation for the introduction (in \cite{H1}) of the generalized
Segal--Bargmann transform for compact Lie groups was the \textquotedblleft$J
$-perp\textquotedblright\ theorem of Leonard Gross \cite{grossErgodic}. That
theorem of Gross is an analog for a compact Lie group $K$ of the Fock space
(symmetric tensor) decomposition on Euclidean space. Gross obtained this
theorem by looking at the pathgroup $W(K)$ along with a Wiener measure $\rho$
on $W(K).$ He then considered the space of functions in $L^{2}(W(K),\rho)$
that are invariant under the left action of the finite-energy loop
group---what we will call loop-invariant functions.

As expected, the space of loop-invariant functions turns out to consist
entirely of functions of the endpoint, but proving this is no small task.
Gross first linearized the problem by mapping paths in the group $K$ to paths
in the Lie algebra $\mathfrak{k}$ by means of the It\^{o} map. The loop
invariant functions in $W(K)$ correspond to functions on $W(\mathfrak{k})$
that are invariant under a certain action of the loop group, which we will
also call loop invariant functions. Gross then expands a loop-invariant
function on $W(\mathfrak{k})$ in a \textquotedblleft chaos
expansion,\textquotedblright\ the infinite-dimensional linear version of the
Fock space decomposition. The Fock space decomposition for the
infinite-dimensional linear space $W(\mathfrak{k}),$ when restricted to
loop-invariant functions gives rise to the $J$-perp expansion for the compact
group $K.$ This analysis also leads to a proof that the only loop-invariant
functions are endpoint functions, that is, functions of the endpoint of the
It\^{o} map. See \cite{HS,DH1} for further results in this direction, and
\cite{ergodic}, \cite{grossCosa}, and \cite{bull} for additional exposition.

The existence of an analog of the Fock space decomposition for $K$ led Gross
to suggest that I look for an analog for $K$ of the Segal--Bargmann transform.
My work on that subject became my Ph.D. thesis and led to the paper \cite{H1}.
Although the motivation for this work was in stochastic analysis, the paper
\cite{H1} was purely finite dimensional. Later on, Gross and Paul Malliavin
showed \cite{GM} that the Segal--Bargmann transform for $K $ could be
understood in much the same way as the $J$-perp expansion. Roughly speaking,
the main result of \cite{GM} asserts that the Segal--Bargmann transform for
$K$ coincides with the Segal--Bargmann transform for the infinite-dimensional
linear space $W(\mathfrak{k}),$ when restricted to functions of the endpoint
of the It\^{o} map. This result is in the vein of much of the work of
Malliavin: using infinite-dimensional analysis to obtain results in
finite-dimensional analysis.

\subsection{A two-parameter version of the Gross--Malliavin result}

In the above discussion of the work of Gross and Malliavin, I have glossed
over the distinction between different forms of the Segal--Bargmann transform
for $K.$ The paper \cite{GM} actually deals with the $B_{t}$ form of the
transform (Theorem $1^{\prime}$ in \cite{H1}), which is nothing but the $s=t$
case of the transform $B_{s,t}$ discussed in the proof of Theorem
\ref{ss.thm}. Driver and I generalize the work of Gross and Malliavin to work
for the transform $B_{s,t}$ and then obtain the transform $C_{t}$ as the
$s\rightarrow\infty$ limit of $B_{s,t}.$ This work was motivated in part by
the work of Gross and Malliavin and in part by the work of K. Wren \cite{Wr}
on the quantization of $(1+1)$-dimensional Yang--Mills theory.

We now examine the details of the construction in \cite{DH1}, with some small
notational changes. Let $H(\mathfrak{k})$ denote the space of absolutely
continuous paths $B:[0,1]\rightarrow\mathfrak{k}$ having one (distributional)
derivative in $L^{2}$ and satisfying $B_{0}=0.$ Let $W(\mathfrak{k})$ denote
the space of continuous paths $B:[0,1]\rightarrow\mathfrak{k}$ satisfying
$B_{0}=0,$ so that $H(\mathfrak{k})$ is a dense subspace of $W(\mathfrak{k}).$
Let $P_{s}$ denote the Wiener measure of variance $s$ on $W(\mathfrak{k}),$
which is characterized by the property that%
\[
\int_{W(\mathfrak{k})}e^{i\phi(B)}dP_{s}(B)=\exp\left(  -\frac{s}{2}\left\Vert
\phi\right\Vert _{H(\mathfrak{k})}^{2}\right)
\]
for all continuous linear functionals $\phi$ on $W(\mathfrak{k}),$ where
$\left\Vert \phi\right\Vert _{H(\mathfrak{k})}^{{}}$ denotes the norm of
$\phi$ as a linear functional on $H(\mathfrak{k}).$ We let $W(K)$ denote the
set of continuous maps $x:[0,1]\rightarrow K$ satisfying $x_{0}=e$ and we let
$\theta:W(\mathfrak{k})\rightarrow W(K)$ denote the It\^{o} map. The It\^{o}
map is the almost-everywhere-defined map sending $B\in W(\mathfrak{k})$ to
$x\in W(K)$ given by solving the Stratonovich stochastic differential equation%
\[
dx_{t}=x_{t}\circ dB_{t}.
\]

We now let $H(\mathfrak{k}_{\mathbb{C}})$, $W(\mathfrak{k}_{\mathbb{C}}),$ and
$W(K_{\mathbb{C}})$ denote the analogously defined spaces with values in the
complexified group or Lie algebra. We let $M_{s,t}$ denote the Wiener measure
on $W(\mathfrak{k}_{\mathbb{C}})=W(\mathfrak{k})\oplus W(\mathfrak{k})$ with
variance $(s-t/2)$ in the real directions and variance $t/2$ in the imaginary
directions, which is nothing but the product measure
\[
dM_{s,t}(A,B)=dP_{s-t/2}(A)~dP_{t/2}(B).
\]
We also have the complex version of the It\^{o} map, $\theta_{\mathbb{C}%
}:W(\mathfrak{k}_{\mathbb{C}})\rightarrow W(K_{\mathbb{C}})$ given by solving
the Stratonovich differential equation%
\[
dg_{t}=g_{t}\circ dZ_{t}.
\]

For positive real numbers $s$ and $t$ with $s>t/2,$ we now have a
Segal--Bargmann transform $S_{s,t}:L^{2}(W(\mathfrak{k}),P_{s})\rightarrow
\mathcal{H}L^{2}(W(\mathfrak{k}_{\mathbb{C}}),M_{s,t})$ given formally by
\[
S_{s,t}(f)=\text{analytic continuation of }e^{t\Delta/2}f,
\]
where $\Delta$ is supposed to represent the sum of squares of derivatives with
respect to an orthonormal basis for $H(\mathfrak{k})$ and the analytic
continuation is from $W(\mathfrak{k})$ to $W(\mathfrak{k}_{\mathbb{C}})$ with
$t$ fixed. The above description of $S_{s,t}$ may be taken more or less
literally on polynomial cylinder functions and the map $S_{s,t}$ then extends
to a unitary map of $L^{2}(W(\mathfrak{k}),P_{s})$ onto $\mathcal{H}%
L^{2}(W(\mathfrak{k}_{\mathbb{C}}),M_{s,t}).$ Here $\mathcal{H}L^{2}$ is
defined as the $L^{2}$ closure of the space of holomorphic polynomial cylinder
functions. I refer to Section 4 of \cite{DH1} for details.

Given a function $f$ on $K,$ we can form the \textquotedblleft endpoint
function\textquotedblright\ on $W(\mathfrak{k})$ given by $B\longmapsto
f(\theta(B)_{1}).$ The endpoint of the It\^{o} map, $\theta(B)_{1},$ is
distributed as the heat kernel measure $\rho_{s}(x)~dx$ on $K,$ which means
that the norm of $f(\theta(B)_{1})$ in $L^{2}(W(\mathfrak{k}),P_{s})$ is equal
to the norm of $f$ in $L^{2}(K,\rho_{s}).$ Similarly, the endpoint of the
complex It\^{o} map is distributed as $\mu_{s,t}(g)~dg,$ so that the norm of
$F(\theta_{\mathbb{C}}(Z)_{1})$ in $L^{2}(W(\mathfrak{k}_{\mathbb{C}}%
),M_{s,t})$ is equal to the norm of $F$ in $L^{2}(K_{\mathbb{C}},\mu_{s,t}).$
Furthermore, $F\in L^{2}(K_{\mathbb{C}},\mu_{s,t})$ belongs to the holomorphic
subspace if and only if $F(\theta_{\mathbb{C}}(\cdot)_{1})\in L^{2}%
(W(\mathfrak{k}_{\mathbb{C}}),M_{s,t})$ belongs to the holomorphic subspace.

\begin{theorem}
\label{dh.thm}Given $f\in L^{2}(K,\rho_{s}),$ let $F=B_{s,t}f,$ which means
that $F$ is the analytic continuation to $K_{\mathbb{C}}$ of $e^{t\Delta/2}f.$
Consider the endpoint function $f(\theta(B)_{1})\in L^{2}(W(\mathfrak{k}%
),P_{s}).$ Then%
\[
S_{s,t}(f(\theta(\cdot)_{1})=F(\theta_{\mathbb{C}}(\cdot)_{1}).
\]

\end{theorem}

That is to say, the Segal--Bargmann transform $S_{s,t}$ for the
infinite-dimensional linear space $W(\mathfrak{k}),$ when restricted to
endpoint functions, becomes the Segal--Bargmann transform $B_{s,t}$ for the
compact group $K.$ This result is Theorem 5.2 of \cite{DH1}. The case $s=t$ is
a variant of one of the main results of \cite{GM}.

Note that the $C_{t}$ version of the Segal--Bargmann transform does not make
sense in the infinite-dimensional linear case. This is because in the $C_{t}$
version, the measure on the domain space should be Riemannian volume measure,
which would be a Lebesgue measure in the linear case, and there is no Lebesgue
measure when the dimension of the space is infinite. Driver and I introduced
the two-parameter transform $S_{s,t}$ with the idea that the large $s$ limit
of this transform would be an approximation to the nonexistent transform
$C_{t}.$

Note that the formula for $B_{s,t}$ is independent of $s$; only the norms on
the domain and range depend on $s.$ Furthermore, as on any compact manifold,
the heat kernel measure $\rho_{s}(x)~dx$ tends to a constant multiple of the
Riemannian volume measure $dx$ as $s$ tends to infinity. Thus, roughly
speaking, the $C_{t}$ form of the Segal--Bargmann transform can be obtained
from the infinite-dimensional linear case by applying $S_{s,t}$ to endpoint
functions and then letting $s$ tend to infinity.

\subsection{Toeplitz operators on $K_{\mathbb{C}}$ from the
infinite-dimensional perspective\label{inftoep.sec}}

We now (finally) arrive at the matter of Toeplitz operators. If $\phi$ is a
bounded measurable function on $W(\mathfrak{k}_{\mathbb{C}})$ then we can
define the Toeplitz operator $T_{\phi}$ on $\mathcal{H}L^{2}(W(\mathfrak{k}%
_{\mathbb{C}}),M_{s,t})$ precisely as in the finite-dimensional case as
$T_{\phi}(\Psi)=P_{s,t}(\phi\Psi),$ where $P_{s,t}$ is the orthogonal
projection from $L^{2}(W(\mathfrak{k}_{\mathbb{C}}),M_{s,t})$ to the
holomorphic subspace. (Recall that the holomorphic subspace is defined to be
the $L^{2}$ closure of the space of holomorphic polynomial cylinder
functions.) For $\Psi_{1},\Psi_{2}\in\mathcal{H}L^{2}(W(\mathfrak{k}%
_{\mathbb{C}}),M_{s,t}),$ we write (as in (\ref{matrix.entries}))
\begin{equation}
\left\langle \Psi_{1},T_{\phi}\Psi_{2}\right\rangle =\int_{W(\mathfrak{k}%
_{\mathbb{C}})}\overline{\Psi_{1}(Z)}\phi(Z)\Psi_{2}(Z)~dM_{s,t}%
(Z).\label{infToep}%
\end{equation}

Up to now, things have been rigorous, but the time has come to shift to a
heuristic viewpoint. It is certainly possible that all of what is to come
could be done rigorously, but for now we content ourselves with using a
heuristic infinite-dimensional argument to give additional insight into the
rigorous finite-dimensional proofs of the preceding section.

Let us now consider how Toeplitz operators relate to multiplication operators.
Suppose $U$ is a function on $W(\mathfrak{k})$ and $U$ is of the form
$U=e^{t\Delta/4}\tilde{U}$ for some other function $\tilde{U}$ (assuming we
can make sense of $e^{t\Delta/4}$). Now, given functions $\psi_{1}$ and
$\psi_{2}$ on $W(\mathfrak{k}),$ we let $\Psi_{1}$ and $\Psi_{2} $ denote the
analytic continuations of $e^{t\Delta/2}\psi_{1}$ and $e^{t\Delta/2}\psi_{2},$
respectively. We use the $s\rightarrow\infty$ limits of the Hilbert spaces
$L^{2}(W(\mathfrak{k}),P_{s})$ and $\mathcal{H}L^{2}(W(\mathfrak{k}%
_{\mathbb{C}}),M_{s,t})$ and the map $S_{s,t}$ as approximations to the
nonexistent spaces $L^{2}(W(\mathfrak{k}),\mathcal{D}A),$ $\mathcal{H}%
L^{2}(W(\mathfrak{k}_{\mathbb{C}}),\nu_{t})$ and the nonexistent transform
$C_{t}$ connecting them. In light of (\ref{infToep}) and the
finite-dimensional result (\ref{rn.form}), it is reasonable to expect that we
will have%
\begin{align}
& \lim_{s\rightarrow\infty}\left\langle \psi_{1},U\psi_{2}\right\rangle
_{L^{2}(W(\mathfrak{k}),P_{s})}\nonumber\\
& =\lim_{s\rightarrow\infty}\int_{W(\mathfrak{k}_{\mathbb{C}})}\overline
{\Psi_{1}(A+iB)}\tilde{U}(A)\Psi_{2}(A+iB)~dM_{s,t}(A,B).\label{inf.lim}%
\end{align}

Now, there are at least two reasons why (\ref{inf.lim}) does not make sense in
general. First, the heat operator $e^{t\Delta/4}$ is very ill behaved in the
infinite-dimensional case. Second, we have to regard the \textit{same}
functions $\psi_{1}$ and $\psi_{2}$ as functions in various \textit{different}
$L^{2}$ spaces of the form $L^{2}(W(\mathfrak{k}),P_{s}).$ However, for $s\neq
s^{\prime},$ the measures $P_{s}$ and $P_{s^{\prime}}$ are mutually singular,
so it does not make sense to think of an element of $L^{2}(W(\mathfrak{k}%
),P_{s})$ as also being an element of $L^{2}(W(\mathfrak{k}),P_{s^{\prime}}).$
A similar issue applies to the functions $\Psi_{1}$ and $\Psi_{2}.$ As a
result, if we hope to apply (\ref{inf.lim}), we need to restrict to a case
where we can make sense of $e^{t\Delta/4}\tilde{U}$ and where it makes sense
to think of the same function as belonging to various different $L^{2}$
spaces, with respect to pairwise singular measures.

The case we are really interested in is the one in which $\psi_{1},$ $\psi
_{2},$ and $U$ are all functions of the endpoint of the It\^{o} map. So we
assume $\psi_{j}(A)=f_{j}(\theta(A)_{1}),$ $j=1,2,$ where $f_{1}$ and $f_{2}$
are functions on $K$ and we assume that $U(A)=V(\theta(A)_{1}).$ If $A$ is
distributed as the measure $P_{s}$ then $\theta(A)_{1}$ is distributed as the
heat kernel measure $\rho_{s}(x)~dx$ on $K.$ These heat kernel measures are
equivalent for different values of $s.$ Thus, for $f$ an almost-everywhere
defined function on $K,$ it makes sense to think of $f(\theta(A)_{1})$ as a
function in various different spaces $L^{2}(W(\mathfrak{k}),P_{s}).$

We compute $\Psi_{1}$ and $\Psi_{2}$ using the transform $S_{s,t}.$ By Theorem
\ref{dh.thm}, we have $\Psi_{j}(Z)=F_{j}(\theta_{\mathbb{C}}(Z)),$ where
$F_{j}$ is the analytic continuation to $K_{\mathbb{C}}$ of $e^{t\Delta
/2}f_{j}.$ Now, Theorem \ref{dh.thm} is one rigorous way of interpreting the
heuristic formula%
\begin{equation}
\Delta\lbrack f(\theta(\cdot)_{1})]=(\Delta_{K}f)(\theta(\cdot)_{1}%
).\label{intertwine}%
\end{equation}
Here on the left-hand side $\Delta$ refers to the infinite-dimensional
Laplacian for $W(\mathfrak{k})$ (formally, sum of squares of derivatives with
respect to an orthonormal basis of $H(\mathfrak{k})$) and on the right-hand
side, $\Delta_{K}$ refers to the finite-dimensional Laplacian for $K.$ (See
also the appendix of \cite{DH1} for another way of interpreting this formula.)
Suppose that there exists a function $\tilde{V}$ on $K$ such that
$V=e^{t\Delta/4}\tilde{V}.$ Then (\ref{intertwine}) implies, at least
formally, that%
\[
e^{t\Delta/4}(\tilde{V}(\theta(\cdot)_{1})=V(\theta(\cdot)_{1}).
\]

We see, then, that in the case of endpoint functions, we can make sense of
both sides of (\ref{inf.lim}). Of course, this does not prove that the two
sides are equal, but it seems reasonable to expect this to be the case. In the
case of endpoint functions, (\ref{inf.lim}) becomes%
\begin{align*}
& \lim_{s\rightarrow\infty}\int_{W(\mathfrak{k})}\overline{f_{1}(\theta
(A)_{1})}V(\theta(A)_{1})f_{2}(\theta(A)_{1})~dP_{s}(A)\\
& =\lim_{s\rightarrow\infty}\int_{W(\mathfrak{k}_{\mathbb{C}})}\overline
{F_{1}(\theta_{\mathbb{C}}(A+iB)_{1})}F_{2}(\theta_{\mathbb{C}}(A+iB)_{1}%
)\tilde{V}(\theta(A)_{1})~dM_{s,t}(A,B).
\end{align*}
We can write this as%
\begin{equation}
\lim_{s\rightarrow\infty}\int_{K}\overline{f_{1}(x)}V(x)f_{2}(x)\rho
_{s}(x)~dx=\lim_{s\rightarrow\infty}\int_{K_{\mathbb{C}}}\overline{F_{1}%
(g)}F_{2}(g)~d\mu_{s,t}^{\tilde{V}}(g),\label{s.toep}%
\end{equation}
where $d\mu_{s,t}^{\tilde{V}}$ is (in general complex) measure defined by%
\[
d\mu_{s,t}^{\tilde{V}}=E_{\ast}\left(  \tilde{V}(\theta(A))~dM_{s,t}%
(A,B)\right)  ,
\]
where $E_{\ast}$ is the push-forward under the \textquotedblleft endpoint
map\textquotedblright\ $E:W(\mathfrak{k}_{\mathbb{C}})\rightarrow
K_{\mathbb{C}}$\ given by%
\[
E(Z)=\theta_{\mathbb{C}}(Z)_{1}.
\]

Let us now assume that $\mu_{s,t}^{\tilde{V}}$ has a density $\mu
_{s,t}^{\tilde{V}}(g)$ with respect to the Haar measure $dg.$ Since $\rho_{s}$
tends to $1/\mathrm{Vol}(K)$ as $s$ tends to infinity, letting $s$ tend to
infinity in (\ref{s.toep}) gives%
\begin{align}
& \frac{1}{\mathrm{Vol}(K)}\int_{K}\overline{f_{1}(x)}V(x)f_{2}%
(x)~dx\nonumber\\
& =\int_{K_{\mathbb{C}}}\overline{F_{1}(g)}F_{2}(g)\left[  \lim_{s\rightarrow
\infty}\frac{\mu_{s,t}^{\tilde{V}}(g)}{\nu_{t}(g)}\right]  \nu_{t}%
(g)~dg.\label{inf.lim2}%
\end{align}
If all of this heuristic arguing actually leads in the end to the right
answer, (\ref{inf.lim2}) tells us that $C_{t}M_{V}C_{t}^{-1}$ can be
represented as a Toeplitz operator with symbol $\phi_{V}$ given by%
\begin{equation}
\phi_{V}(g)=\mathrm{Vol}(K)\lim_{s\rightarrow\infty}\frac{\mu_{s,t}^{\tilde
{V}}(g)}{\nu_{t}(g)}.\label{phiv.inf}%
\end{equation}

The infinite-dimensional approach thus at least gives us a prediction of what
the Toeplitz symbol of $C_{t}M_{V}C_{t}^{-1}$ should be. We will now verify
(rigorously) that the right-hand side of (\ref{phiv.inf}) agrees with the
expression for $\phi_{V}$ given in Section \ref{mult.sec}. We restrict
ourselves to the semisimple case; the commutative case is similar. I am
grateful to Bruce Driver for pointing out to me the relation (\ref{theta}) and
its proof.

We begin by observing that%

\begin{equation}
\theta_{\mathbb{C}}(A+iB)=\theta_{\mathbb{C}}(iB^{\theta(A)})\theta
(A),\label{theta}%
\end{equation}
almost surely, where%
\[
B_{t}^{\theta(A)}=\int_{0}^{t}\mathrm{Ad}_{\theta(A)_{s}}dB_{s}.
\]
This result follows from the stochastic differential equation for
$\theta_{\mathbb{C}}.$ (If $A$ and $B$ were smooth paths, then a simple
computation shows that the right-hand side of (\ref{theta}) would satisfy the
same differential equation as the left-hand side. Stratonovich stochastic
differential equations are such that the same result holds in the stochastic
case, with no correction terms.)

Since the increments of $B$ are distributed in an Ad-$K$-invariant fashion,
the distribution of $(A,B^{\theta(A)})$ is the same as the distribution of
$(A,B).$ Using this fact and (\ref{theta}), we have%
\begin{align}
& \int_{W(\mathfrak{k}_{\mathbb{C}})}f(\theta_{\mathbb{C}}(A+iB))\tilde
{V}(\theta(A))~dM_{s,t}(A,B)\nonumber\\
& =\int_{W(\mathfrak{k}_{\mathbb{C}})}f(\theta_{\mathbb{C}}(iB^{\theta
(A)})\theta(A))\tilde{V}(\theta(A))~dM_{s,t}(A,B)\nonumber\\
& =\int_{W(\mathfrak{k}_{\mathbb{C}})}f(\theta_{\mathbb{C}}(iB)\theta
(A))\tilde{V}(\theta(A))~dM_{s,t}(A,B)\label{pushA}%
\end{align}
for any continuous function $f$ of compact support on $K_{\mathbb{C}}.$ Recall
that $M_{s,t}$ decomposes as the product measure $dP_{s-t/2}(A)\times
dP_{t/2}(B).$ Furthermore, $\theta_{\mathbb{C}}(iB)$ is distributed as the
heat kernel measure $\mu_{s,t}(g)~dg$ on $K_{\mathbb{C}}$ and $\theta(A)$ is
distributed as the heat kernel measure $\rho_{s-t/2}(x)~dx$ on $K.$ Thus,
(\ref{pushA})\ becomes%
\begin{align}
& \int_{W(\mathfrak{k}_{\mathbb{C}})}f(\theta_{\mathbb{C}}(A+iB))\tilde
{V}(\theta(A))~dM_{s,t}(A,B)\nonumber\\
& =\int_{K}\int_{K_{\mathbb{C}}}f(gx)\tilde{V}(x)~\mu_{s,t}(g)~dg~\rho
_{s-t/2}(x)~dx.\label{pushB}%
\end{align}
After making the change of variable $g\rightarrow gx^{-1}$ in the inner
integral of (\ref{pushB}) and reversing the order of integration, we see that
the pushed-forward measure (which we are denoting $d\mu_{s,t}^{\tilde{V}}(g)$)
is given by%

\begin{equation}
d\mu_{s,t}^{\tilde{V}}(g)=\left[  \int_{K}\mu_{s,t}(gx^{-1})\tilde{V}%
(x)\rho_{s-t/2}(x)~dx\right]  ~dg.\label{push5}%
\end{equation}

If we now let $s$ tend to infinity in (\ref{push5}), $\rho_{s-t/2}$ becomes
$1/\mathrm{Vol}(K)$ and $\mu_{s,t}$ becomes the subelliptic heat kernel
$\mu_{t/2,t}.$ Thus, we have%
\begin{equation}
\mathrm{Vol}(K)\lim_{s\rightarrow\infty}\frac{\mu_{s,t}^{\tilde{V}}(g)}%
{\nu_{t}(g)}=\frac{\int_{K}\mu_{t/2,t}(gx^{-1})\tilde{V}(x)~dx}{\nu_{t}%
(g)},\label{lim}%
\end{equation}
which means that (\ref{phiv.inf}) is in agreement with the results of Section
\ref{mult.sec}.

\end{document}